\documentclass[superscriptaddress,reprint,aps,amsmath,amssymb,prl]{revtex4-1}
\usepackage[caption=false]{subfig}
\usepackage{lipsum}  
\usepackage{lineno}
\usepackage{hyperref}
\usepackage{graphicx}
\usepackage[normalem]{ulem}
\graphicspath{{./figs/}}
\usepackage{amsmath,amssymb}
\hypersetup{
	colorlinks = true,
	urlcolor   = blue,
	citecolor  = blue,
	linkcolor = blue
}
\modulolinenumbers[5]

\newcommand{\Wi}{\mathrm{Wi}}
\renewcommand{\Re}{\mathrm{Re}}

\usepackage{xcolor}

\begin{document}

\preprint{APS/123-QED}

\title{Non-modal elastic instability and elastic waves in weakly perturbed channel flow}

%Elastic waves above elastically driven instability in weakly perturbed channel flow 

\author{Ron Shnapp}
\affiliation{Department of Physics of Complex Systems, Weizmann Institute of Science, Rehovot 76100, Israel}
\author{Victor Steinberg}
\affiliation{Department of Physics of Complex Systems, Weizmann Institute of Science, Rehovot 76100, Israel}
\affiliation{The Racah Institute of Physics, Hebrew University of Jerusalem, Jerusalem 91904, Israel}

\begin{abstract}
	In this paper, we present experimental results and reveal that strong perturbations are not necessary for elastic instability to occur in straight-channel, inertialess, visco-elastic flows at high elasticity. We show that a non-normal mode bifurcation is followed by chaotic fluctuations, self-organized as stream-wise streaks, and elastic waves due to weak disturbances generated by a small cavity at the center of the top channel wall. The chaotic flow persists in the transition, elastic turbulence, and drag reduction regimes, in agreement with previous observations for the case of strong perturbations at the inlet. Furthermore, the elastic waves we observe propagate in the span-wise direction, which allows to confirm the elastic waves linear dispersion relation directly for the first time. In addition, the span-wise propagating elastic wave's velocity depends on $\mathrm{Wi}$ with the same scaling that was previously observed for stream-wise propagating waves, although their velocity magnitude is significantly smaller than what was previously observed for the stream-wise ones.
\end{abstract}

\pacs{pacs1}% PACS, the Physics and Astronomy
% Classification Scheme.
%Use show keys class option if keyword
%display desired
\maketitle

\section{introduction}

Dissolving tiny amounts of long, flexible, and linear polymer molecules in viscous Newtonian solvents strongly affects the dynamics of both laminar and turbulent flows due to polymer stretching by flow~\cite{Bird1987}. As a result, even in low-velocity shear flows, such as the one discussed in this paper, significant elastic stress can be generated due to large velocity gradients and high polymer longest relaxation times. Indeed, elastically driven instabilities~\cite{Larson1992, Shaqfeh1996} and a chaotic flow state called elastic turbulence (ET)~\cite{Groisman2000} are observed in shear flows at low Reynolds number, $\Re\equiv U L \rho/\eta \ll 1$, and high Weissenberg number, $\Wi\equiv U/ L \lambda \gg 1$, corresponding to high fluid elasticity, $\mathrm{El}\equiv \Wi/\Re\gg 1$~\cite{Steinberg2021}. Here, $U$, $L$, $\rho$, and $\eta$ are the characteristic velocity and length scales, the fluid's density and viscosity, respectively, and $\lambda$ is the longest polymer relaxation time. $\Wi$ is a control parameter that defines the degree of polymer stretching by the flow~\cite{Steinberg2021}.

In inertialess flows ($\Re \ll 1$) with curvilinear streamlines, the elastic stress that develops in polymeric fluids destabilizes the flow, causing ET and effective mixing~\cite{Groisman2001, Groisman2000}. These flows are linearly unstable, so the most unstable normal mode grows exponentially and then saturates at sufficiently large amplitude due to nonlinear interactions~\cite{Larson1990, Shaqfeh1996, Drazin2004}. In such shear flow geometries, the elastic instability is driven by elastic stress along the curved streamlines, which initiates a force in the curvature direction, causing the instability \cite{Larson1990, Pakdel1996, Shaqfeh1996}. However, this mechanism is ineffective in parallel-shear flows with zero curvature, such as the Poiseuille and plane Couette flows, and their linear stability has been proved~\cite{Gorodtsov1967, Renardy1986, Morozov2007}.

The linear stability of parallel-shear flows does not imply they are globally stable. Two well-known and recently investigated examples are the pipe, and the plane Couette flows of Newtonian fluids, which become unstable at finite $\Re$ even though they are linearly stable for all $\Re$~\cite{Reynolds1883, Avila2011}. To explain the observation of such instabilities in a broad class of parallel-shear Newtonian flows, a new concept, called the non-normal mode instability, was introduced as an alternative to the traditionally accepted least-stable normal Eigen-mode bifurcation~\cite{Schmid2007, Trefethen1993}. The new approach is based on the fact that the Orr-Sommerfeld equation, which describes the linear instability of parallel-shear flows~\cite{Drazin2004}, is a non-self-adjoint operator~\cite{Trefethen1993}. Notably, while flows with normal-mode instability are sensitive to even infinitesimally small perturbations, the non-normal instability requires a finite-sized perturbation to develop. Then, in addition to the stable normal modes, unstable non-orthogonal modes emerge and grow algebraically up to a sufficiently large amplitude due to finite-size perturbations. A theory for non-modal instability in visco-elastic channel and pipe flows for the $El\gg1$ regime was introduced in Refs.~\cite{Jovanovic2010, Jovanovic2011, Lieu2013}. The resulting span-wise modulated coherent structures were observed in both Newtonian~\cite{Grossmann2000, Schoppa2002} and visco-elastic~\cite{Jha2021, Jha2020} channel flows. 

Several works have experimentally shown the elastic instability that occurs in inertialess straight-shear flows. In particular, Ref.~\cite{Bonn2011} first reported strong velocity fluctuations in a pipe flow, where a strong jet was used to perturb the fluid at the pipe's inlet. Furthermore, experiments in a square micro-channel~\cite{Pan2013, Qin2017, Qin2019} demonstrated that strong prearranged perturbations by a set of obstacles at the inlet of a channel flow lead to strong flow fluctuations above the instability threshold value, $\Wi_c$. A recent experiment from our lab studied the elastic instability and characterized the flow in a quasi-2D channel~\cite{Jha2020, Jha2021}. In this experiment too, the flow was strongly perturbed by placing an array of obstacles at the inlet covering the full channel width. Three flow regimes were observed in this experiment above the instability onset: transition, ET, and drag reduction (DR), where the bifurcation was found to be continuous and non-hysteretic. Moreover, as determined by examining three different features, it was observed that the instability was a non-normal mode bifurcation. First, the $\Wi$ dependence of the normalized friction factor, $C_f/C_f^{\mathrm{lam}}$, and of the root mean squared (RMS) velocity and pressure fluctuations, $u_{\mathrm{RMS}}$ and $P_{\mathrm{RMS}}$, have slope exponents that are significantly different from the normal-mode bifurcation indicative value of 0.5. Second, just above the instability onset, there was a continuous velocity power spectrum with an algebraic decay at higher frequencies, in addition to high energy peaks in the span-wise velocity spectrum at lower frequencies~\cite{Jha2021}. Thus, an infinite number of modes are excited above the instability threshold, contrary to the single most unstable mode in the case of the normal mode instability~\cite{Drazin2004}. Third, the $\Wi$ dependence of the spectral peak at low frequency implies the existence of elastic waves on top of a chaotic flow in the transition, ET, and DR regimes. These features confirm that the instability is the non-normal mode bifurcation excited by finite-size perturbations. Furthermore, in all three regimes the flow exhibits weakly unstable coherent structures (CSs) in the form of stream-wise rolls and streaks, organized into a cycling self-sustained process (SSP)~\cite{Waleffe1997, Jha2020}. The CSs selected by the flow depend on the structure of perturbations and can vary for different initial perturbations, as established in the Newtonian flow case~\cite{Trefethen1993, Schmid2007}. This sensitivity to initial conditions is a characteristic property of the non-normal mode bifurcation. Therewith, only in ET, the streaks are destroyed by a secondary instability that strongly resembles the temporal dynamics of the inertial Kelvin-Helmholtz instability, even though the instability mechanism is purely elastic~\cite{Jha2021}. The sequence of CSs repeats itself periodically with the frequency of elastic waves, which pump energy into the cycle and CSs. This characteristic feature of ET distinguishes it from inertial turbulence in parallel shear flows.

The non-normal mode instability of visco-elastic straight channel flows emphasizes the importance of the details of the finite-size flow perturbations. In particular, it raises the question of how sensitive this type of flow is to the initial perturbation's strength and morphology? Therefore, in this paper, we intend to elucidate the following questions: (i) Are the strong prearranged perturbations at the inlet necessary to get elastic instability in straight channel flows of visco-elastic fluids? If not, then (ii) will the flow exhibit the same characteristics of the non-normal mode bifurcation as in Ref.~\cite{Jha2020, Jha2021}? (iii) In this case, will the flow structure be similar to that of the strongly perturbed flow, namely, three flow regimes with similar CSs? (iv) Will elastic waves be observed in this flow as well? And (v) will there be a correlation between the behavior of elastic wave intensity and flow characteristics as a function of $\Wi$ in three flow regimes? To answer the above questions, we conducted an experiment of a visco-elastic fluid straight-channel flow disturbed by a weak perturbation. In particular, unlike previous studies ~\cite{Bonn2011, Pan2013, Qin2017, Qin2019, Jha2021, Jha2020}, the inlet to our channel was carefully smoothed and tapered and without any obstacles blocking the flow. We find that this detail is crucial, since wall roughness and imperfections at the inlet may cause an earlier elastic instability, similar, e.g., to Newtonian pipe flow in which the stronger the perturbations, the lower the instability onset \cite{Hof2003}. Instead, the only flow perturbations are generated by a small cavity at the channel's top wall, close to the middle of the channel length. The disturbance to the channel flow was thus initiated by velocity fluctuations caused by an elastic instability and ET inside the cavity, driven by a stream-wise channel flow at the boundary. In particular, the flow inside the cavity has curved streamlines and is therefore subjected to a normal bifurcation and exhibits ET at higher $\Wi$. Such flow geometry was investigated earlier in Ref.~\cite{Pakdel1998}, where an experimental study of the transition in a lid-driven square cavity flow with curved streamlines was presented~\cite{Shaqfeh1996, Pakdel1996}. Notably, the cavity flow in this study was driven by a moving solid boundary. However, later on, the same elastic instability was investigated in a wide square cavity connected to a channel ~\cite{Kim2000}. Thus, the perturbations and flow instability downstream from the cavity was triggered by the flow inside the cavity above the elastic instability, similar to the instabilities observed in ~\cite{Pakdel1998, Kim2000}, and further on, in ET.

Similar to the strongly perturbed channel flow \cite{Jha2020}, in this paper, we study the flow structure and properties at several locations downstream from the cavity, since upstream from it, the flow is found to be laminar. The key message of the paper is that a weak, finite-size perturbation, due to the presence of a cavity in a straight channel shear flow, is sufficient to cause an elastic instability along with the transition, ET, and DR regimes, where stream-wise velocity streaks are observed in all of them. Another striking observation is span-wise elastic waves with the same power-law dependence on $\Wi$, but with a much lower velocity than the stream-wise propagating elastic waves shown in Refs.~\cite{Varshney2019, Jha2020}.

\begin{figure*}[!htb]
	\centering
	\includegraphics[height=5.2cm]{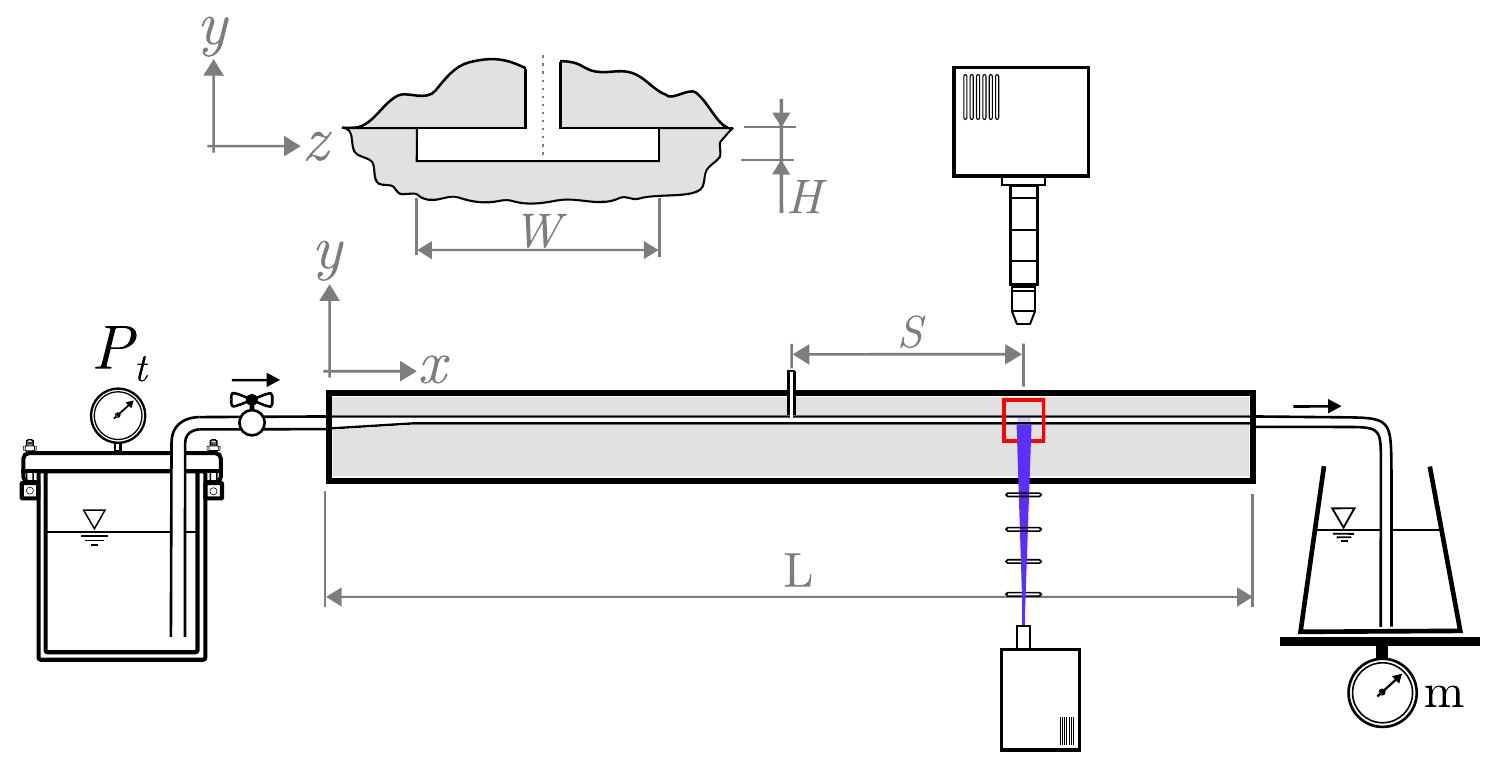}
	\caption{Schematics of the channel flow apparatus and the PIV system used in the experiments, along with a cross-section view at the center of the channel showing the cavity of diameter d=H. The channel dimensions are length (L) $\times$ width(W) $\times$ height(H)=750$\times$3.5$\times$0.5 $mm^3$. The horizontal $x-z$ mid-plane is located at $y=0$. \label{fig:setup}}
\end{figure*}

\begin{figure}
	\centering
	\includegraphics[width=8.5cm]{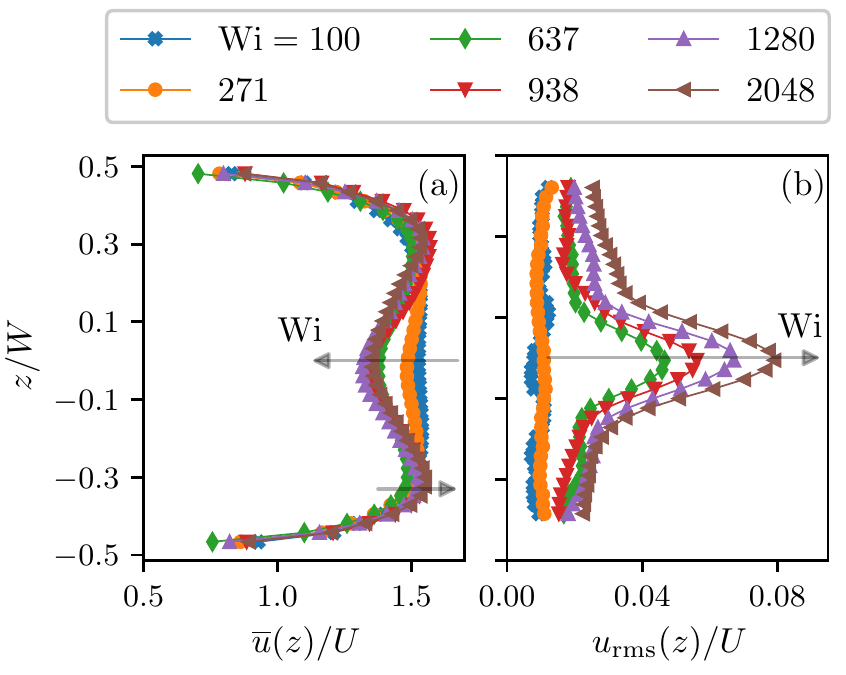}
	\caption{(a) Mean velocity, and (b) root mean square of the stream-wise velocity, plotted as a function of the span-wise coordinate and for several $\Wi$ numbers. Data taken at a distance $S=40H$ downstream from the cavity. \label{fig:profiles}}
\end{figure}

\begin{figure}
	\centering
	\includegraphics[width=8.58cm]{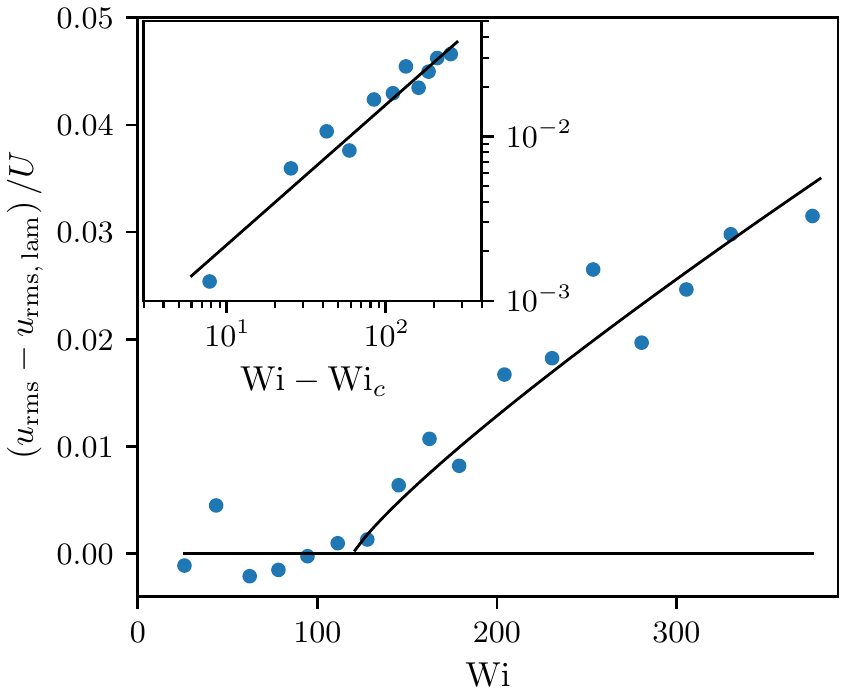}
	\caption{The normalized RMS of stream-wise velocity fluctuations at the center of the channel, $z=0$ and $S=71H$, plotted as a function of $\Wi$. The main panel and the inset show the same data in linear scale and in log-log scales respectively. The experimental noise level was subtracted from the data series by a linear fit of the data before the transition. The fit of $u_{\mathrm{RMS}}\propto (\Wi-\Wi_c)^a$ for $\Wi>\Wi_c$ is shown as black lines in both pannels with values of $\Wi_c = 120\pm15$ and $a=0.85\pm0.15$. \label{fig:uRMS_transition}}
\end{figure}

\begin{figure}[!htb]
	\centering
	\subfloat{
		\includegraphics[width=8.58cm]{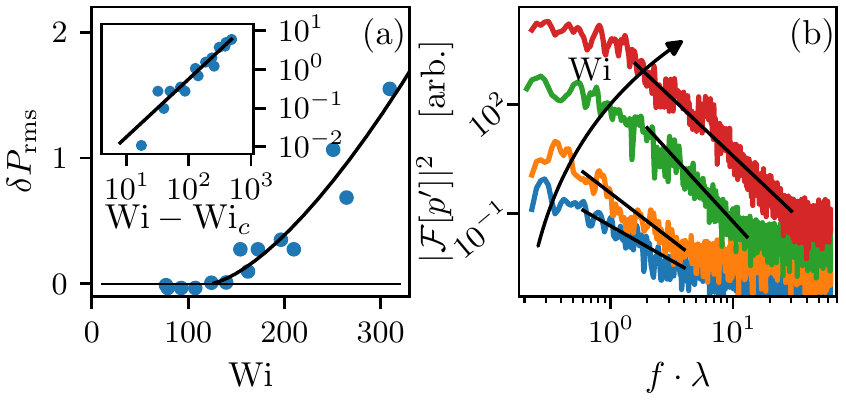}
	}
	\caption{(a) stream-wiseof pressure fluctuations, normalized by experimental noise measured at the cavity location. A fit of the data (black line) gives $\delta P_{\mathrm{RMS}} \propto (\Wi - \Wi_c)^{\alpha}$ with $\alpha=1.5\pm 0.2$. The inset shows the same data in log-log scales as a function of $\Wi-\Wi_c$ to better present the scaling. (b) Power spectra of the pressure fluctuations for $\Wi=$32, 264, 760, and 1478, shown in log-log scales along with straight lines which represent scaling exponents of -1.9, -2.6, -3.7 and -3.2, respectively. The curves were translated vertically for better visualization. \label{fig:dp_RMS}}
\end{figure}

% ===============================
% ========== Methods ============
% ===============================

\section{Experimental techniques and methods}

{\it Experimental setup.} The experiment was conducted in a transparent acrylic channel, as shown in Fig.~\ref{fig:setup}. The length, height, and width of the channel are $L=750$ mm, $H=0.5$ mm, and $W=3.5$ mm, respectively. The channel inlet was carefully smoothed and tapered over a distance of roughly $200H$ to eliminate any unwanted flow disturbances. In addition, we drilled a small cavity, which initially served as a port for pressure measurements. Our observations revealed that the cavity triggered downstream of it velocity perturbations larger than those from the inlet. Therefore, separate measurements of flow characteristics, excluding pressure fluctuations, were conducted first with the rather deep cavity of $D=0.5$ mm diameter and about $h=5$ mm in depth, blocked from the pressure sensor tube. Then, pressure fluctuation measurements were made separately, with the end of the cavity connected to the pressure sensor. The fluid in the channel was driven by Nitrogen gas pressurized up to 60 psi.

{\it Preparation and characterization of polymer solution.} As a working fluid, we used an aqueous polymer solution comprised of 44\% Sucrose, 22\% D-Sorbitol, 1\% Sodium Chloride (\textit{Sigma Aldrich}), and $c=230$ ppm Polyacrylamide (PAAm, $\mathrm{Mw}=18\times 10^6 Da$ from \textit{Polysciences inc.}) with $c/c*\simeq 1$, where $c*$ is the overlap polymer concentration~\cite{Liu2009}. The solution density, solvent viscosity and the total viscosity were $\rho = 1320 \mathrm{Kg\,m^{-3}}$, $\eta_s=0.093 \, \mathrm{Pa\, s}$, and $\eta = \eta_p + \eta_s = 0.125 \, \mathrm{Pa\, s}$. In addition, the longest polymer relaxation time for this solution was measured in Ref.~\cite{Liu2009} to be $\lambda=12.1 \,\mathrm{s}$ using the stress relaxation method. 

%Let us note that the relaxation time can be defined experimentally using two methods - either the stress relaxation method or the steady-state shear~\cite{Pakdel1998}. For the sake of future comparison, we give a rough estimate for the steady-state relaxation time,  $\lambda_s$, by employing the measurements of Ref.~\cite{Pakdel1998}. There, measurements for two different solutions resulted in $\lambda_s / \lambda = 0.166$ and $0.188$, so we may estimate that roughly $\lambda_s \approx 2.15\pm0.15$ s for our solution.     

{\it Flow rate and pressure fluctuations measurements.} During each experiment, we used a PC-interfaced balance (BPS-1000-C2-V2, MRC) to measure the time-averaged mass discharge rate, $\Delta m / \Delta t$. Thus we calculated the average velocity in the channel, $U=\frac{\Delta m}{\Delta t} / (\rho \, H \, W)$. Experiments were conducted at low Reynolds numbers $\Re=\frac{\rho U H}{\eta} < 0.7$ and the Weissenberg number was in the range $\Wi=\frac{U}{H} \lambda \in (100, 2100)$. This confirms that our measurements were performed at a high elasticity number, $\mathrm{El}=\frac{\Wi}{\Re}=\frac{\eta\lambda}{\rho H^2}=4583$. We also measured pressure fluctuations using a high-resolution pressure sensor of accuracy $0.1\%$ of a full scale (Honeywell, HSC Series).

{\it Imaging system and high-resolution PIV.} We conducted measurements of the velocity field at various distances, $S$, downstream from the cavity, using the particle image velocimetry (PIV) method. For that, we illuminated small tracer particles (3.2$\mu \mathrm{m}$ fluorescent tracers) with a thin laser sheet (thickness of $\sim30\mu \mathrm{m}$) over the central plane, $y=0$, in the channel. We then captured pairs of images of the particles using a high-speed camera (Photron FASTCAM Mini UX100) with time separations in the range of 8-0.25 ms, depending on the flow rate. The image pairs were recorded at repetition rates of $10-50$ Hz depending on the particular aim of the measurement, where the timing was achieved through an external function generator. The OpenPIV software~\cite{openpiv} was used to calculate 2D velocity field components $u_x(t,x,z)$, $u_z(t,x,z)$. We typically recorded data for periods of $\sim \mathcal{O}(10)$ minutes or $\sim\mathcal{O}(50\lambda)$ for each $\Wi$ to obtain sufficient statistics.

\begin{figure*}
	\centering
	\includegraphics[width=17.16cm]{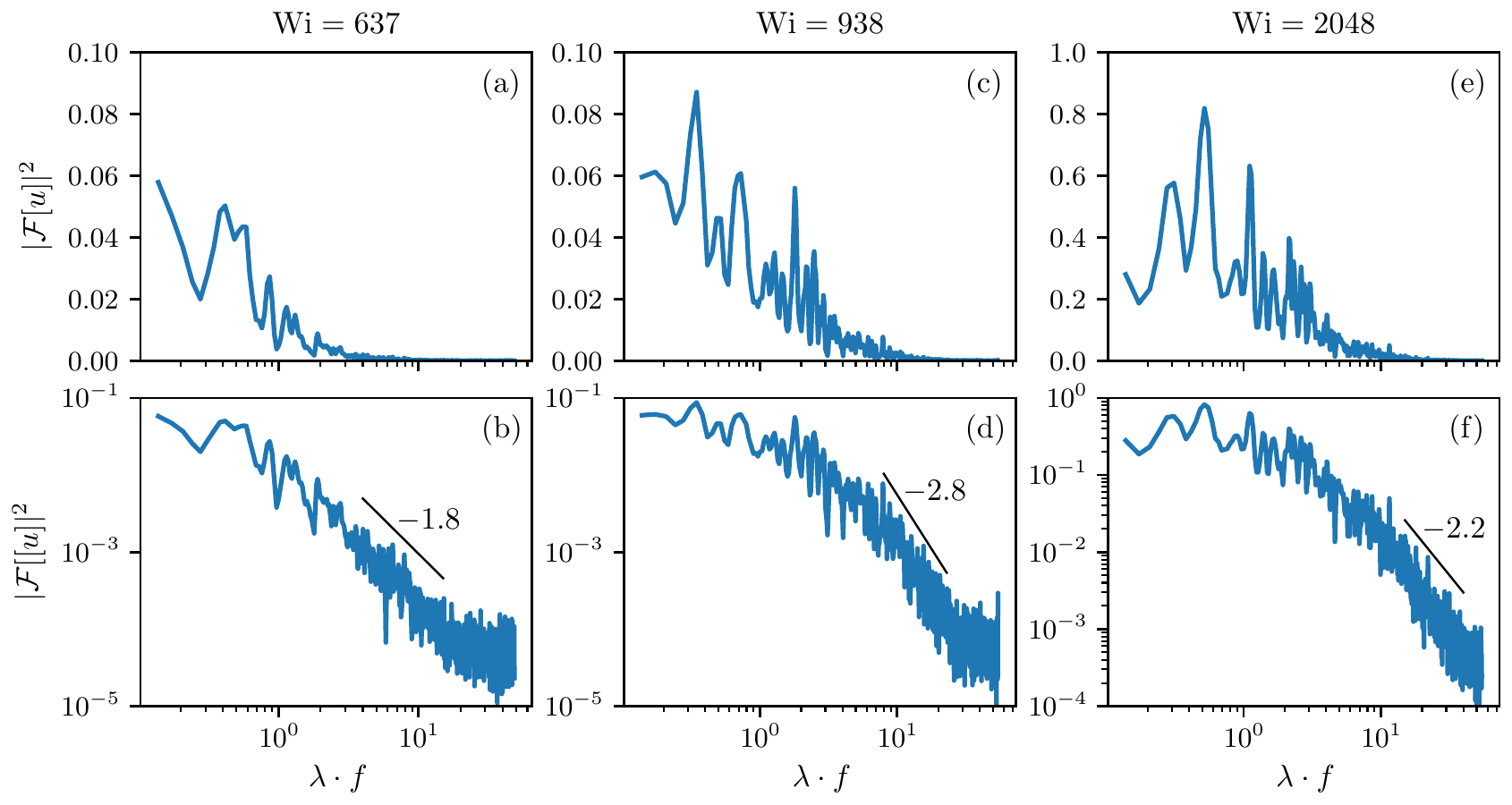}
	\caption{The spectra of stream-wise velocity fluctuations at the center of the channel, $z=0$, shown for three values of $\Wi$ at $S=40H$. The top row shows the spectra in linear-log scales, while the bottom row shows the same data in log-log scales. Data fits to the scaling range of of the data are shown as black lines, where the uncertainty of the exponents is estimated as $\pm 0.1$ \label{fig:40H_spectra}.}
\end{figure*}

\begin{figure}
	\centering
	\includegraphics[width=8.58cm]{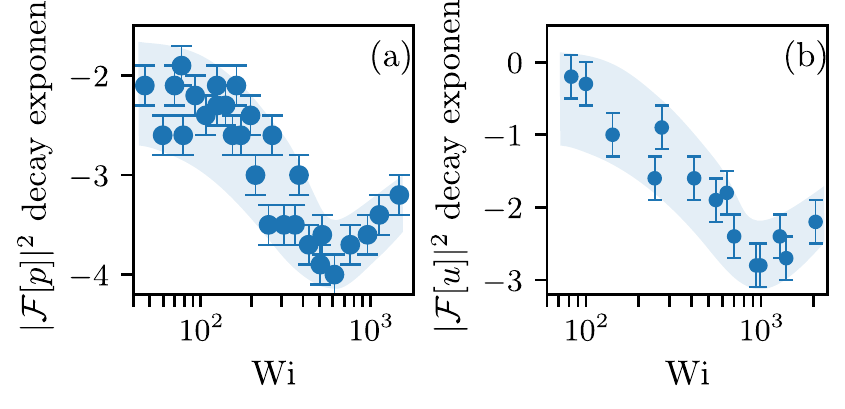}
	\caption{Decay exponent of the spectra of (a) the pressure, and (b) the velocity fluctuations taken at $S=20H$ and $40H$, are shown for various values of $\Wi$. The shaded area, drawn by hand fitting smoothing splines, represents the general trend of the data.}
	\label{fig:spectra_slope}
\end{figure}

% =========================================================
%                          Results
% =========================================================

\section{Results}

{\it Elastic instability, elastic turbulence, and drag reduction.} Once the onset of the elastic instability is crossed, $\Wi>\Wi_c$, and particularly in the ET flow regime inside the cavity, the velocity fluctuations perturb the channel flow, and their effects on a stream-wise channel flow velocity are measured via PIV downstream the cavity. In Fig.~\ref{fig:profiles}a, we show the mean velocity profile in the horizontal mid-plane of the channel flow at a distance $S=40H$ downstream of the cavity. With increasing $\Wi$, a low mean velocity region, a velocity deficit, in the mean velocity profile at the center of the channel ($z=0$) emerges. The velocity deficit, first detected at the critical value for the instability $\Wi \approx \Wi_c$ ($\Wi_c$, is presented in Fig.~\ref{fig:uRMS_transition}). The velocity deficit grows and becomes increasingly more pronounced as $\Wi$ increases in the range of $\Wi_c\lesssim \Wi \lesssim 600$; for $\Wi\gtrsim 600$, the velocity deficit saturates. As a result of the velocity deficit, and even though the cavity does not directly obstruct the flow, the mean velocity profile somewhat resembles that of a flow past an obstacle.

%The flow velocity fluctuates downstream of the cavity for $\Wi>\Wi_c$. 

In Fig. \ref{fig:profiles}b, we show profiles of the stream-wise velocity fluctuations normalized by $U$, $u_{\mathrm{RMS}}/U$, for several values of $\Wi$. For the two lowest values of $\Wi=100$ and 271, we observe weak fluctuations down to $u_{\mathrm{RMS}}\approx 0.01 U$. However, for higher $\Wi$, the velocity fluctuations are significantly stronger and increase at the highest $\Wi$ values up to $8\%$ at the center of the channel. The profile of $u_{\mathrm{RMS}}/U$ versus $z/W$ resembles a Gaussian curve with a total width of roughly $0.4W$ (see Fig.~\ref{fig:profiles}b). 

To better characterize the elastic instability of the channel flow, we present $(u_{\mathrm{RMS}} - u_{\mathrm{RMS,lam}})/U$ at the center of the channel ($z=0$) vs $\Wi$ in Fig.~\ref{fig:uRMS_transition}, taken at $S=71H$ downstream the cavity, where $u_{\mathrm{RMS,lam}}$ is the stream-wise velocity fluctuations recorded due to experimental noise. The fit of an increase of $(u_{\mathrm{RMS}} - u_{\mathrm{RMS,lam}})/U$ versus $\Wi$ above an instability threshold, gives $(u_{\mathrm{RMS}} - u_{\mathrm{RMS,lam}})/U \sim (\Wi - \Wi_c)^a$ with $\Wi_c=120\pm 15$ and $a = 0.85 \pm 0.15$. Notably, $a$ significantly differs from $0.5$, the value expected for the normal-mode bifurcation~\cite{Drazin2004}. Furthermore, the elastic transition is continuous without hysteresis.

Another way to detect and characterize the elastic instability is through the $\Wi$ dependence of the RMS of pressure fluctuations, as shown in Fig.~\ref{fig:dp_RMS}a. Notably, the sharp increase of the pressure fluctuations above the instability threshold is significant, reaching up to 40 times the experimental noise level and up to roughly 35\% of the wall shear stress in the channel ($\approx 0.35 \, \tau_w \equiv 0.35 \, P_{tot} H\,W/2L(H+W)$, where $P_{tot}/L$ is the pressure gradient over the full channel length). Also, no pressure fluctuations above the experimental noise limit were observed for the Newtonian fluid without additional polymers as well as the polymer solution below the elastic instability (see Fig. S1 in Supplementary Materials~\cite{supplamentaryMaterial}). This confirms that the instability is a pure elastic effect that is caused by elastic stress that results from polymer stretching. As shown in Fig.~\ref{fig:dp_RMS}a, the normalized pressure fluctuations, $\delta P_{\mathrm{RMS}}=P_{\mathrm{RMS}}/P_{\mathrm{RMS,lam}}-1$, grow with the Weissenberg number as $\sim (\Wi-\Wi_{c})^\alpha$, where using least square fitting with the same instability threshold obtained above $\Wi_c = 120\pm15$, we obtain the exponent $\alpha = 1.5\pm0.2$. The values of the exponents $a$ and $\alpha$ differ significantly, and they are both significantly higher than $0.5$. Also and as observed above, the elastic transition is continuous and without hysteresis.

\begin{figure*}
	\centering
	\includegraphics[width=17.16cm]{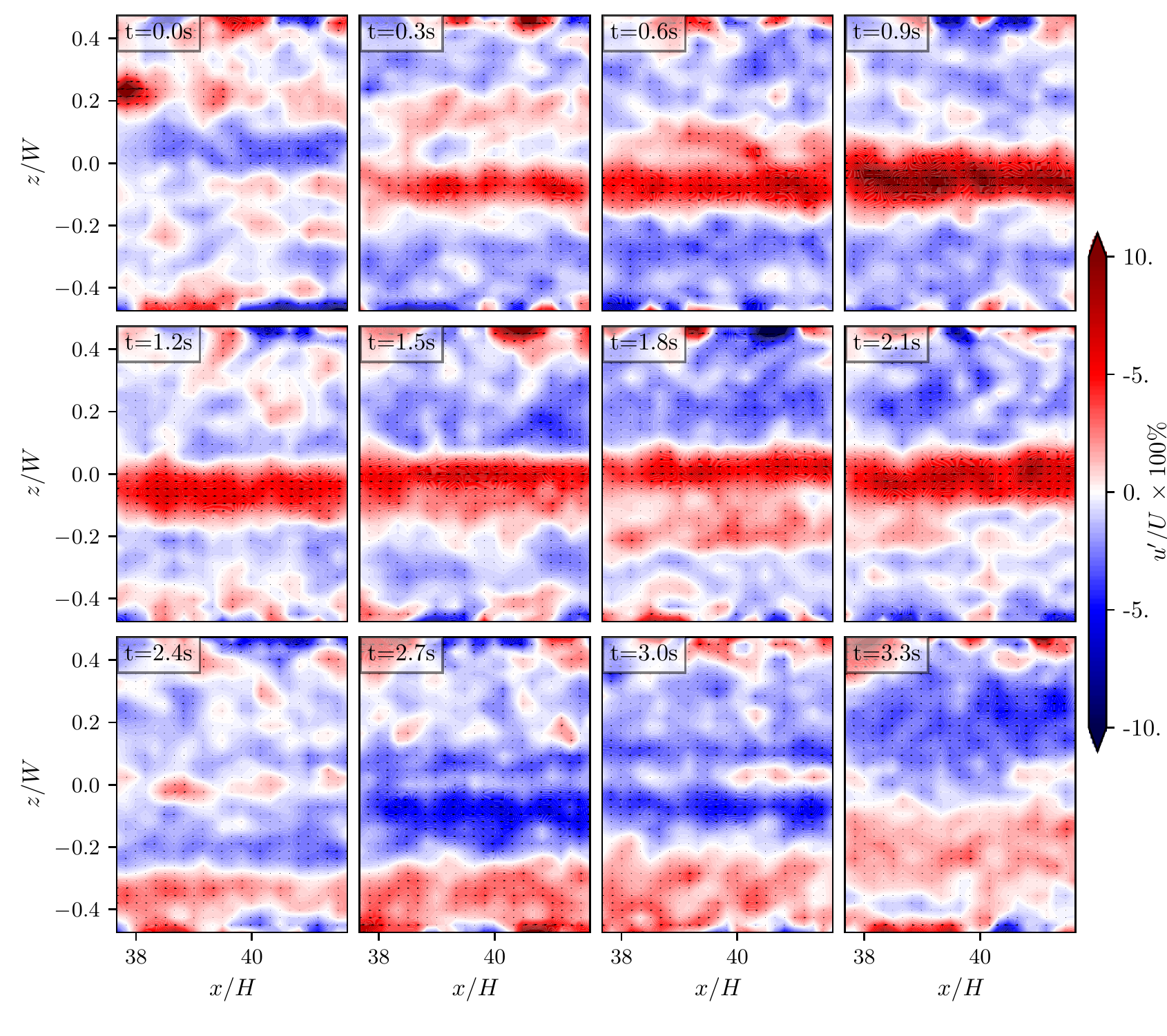}
	\caption{A series of images that demonstrate the time-dependent streaky structure of the flow at $\Wi=938$ taken in ET at $S=40H$. Specifically, each instantaneous image shows the stream-wise velocity fluctuations above its subtracted time-averaged profile as a contour plot, and small arrows show the direction of the velocity fluctuations field. \label{fig:40H_structure}}
\end{figure*}

The stream-wise velocity frequency power spectra, obtained from velocity time series taken at the channel center-line ($z=0$) and $S=40H$, are shown in Fig.~\ref{fig:40H_spectra} for $\Wi=637,$ 938, and 2048. These $\Wi$ values represent the three flow regimes above $\Wi_c$, namely, transition, ET, and DR earlier observed in Ref.~\cite{Jha2020}. The power spectra shown in Fig.~\ref{fig:40H_spectra} are continuous, and they are characterized by an algebraic decay at higher frequencies, $|\mathcal{F}[u]|^2 \propto f^m$, along with the presence of peaks at lower frequencies, typically $f\lesssim\lambda^{-1}$, which are particularly pronounced in lin-log coordinates (Fig.~\ref{fig:40H_spectra}a, c, and e). The typical decay exponents of the spectra, $m$, are shown in Fig.~\ref{fig:40H_spectra}b, d, and f. Similarly, we also calculated the power spectrum of pressure fluctuations, defined as $|\mathcal{F}[p]|^2$, using pressure time series, as demonstrated in Fig.~\ref{fig:dp_RMS}b for $\Wi=32$, 264, 760, and 1478. Here too, at high frequencies the spectra exhibit a power-law decay, $\mathcal{F}\sim f^n$, with the exponent $n$. The values of $n$ for the cases shown in Fig.~\ref{fig:dp_RMS}b that we obtained from the fit are $n=-1.9 \pm 0.2$ at $\Wi<\Wi_c$, $n=-2.6\pm 0.2$ in the transition regime, $-3.7\pm0.2$ in ET, and $-3.2\pm0.2$ in DR, respectively. Notably, the algebraic decay of the pressure spectrum below the instability is attributed in part to experimental $1/f$, "pink" noise in the electrical circuits of the pressure sensor, and in part to the instability inside the cavity itself, as we discuss below in the discussion section.

The values of the decay exponents for both the pressure and velocity power spectra, $n$ and $m$, are presented as a function of $\Wi$ in Fig.~\ref{fig:spectra_slope}. Despite the scatter in the results that is associated with experimental noise, a trend can be clearly detected in both cases. For the pressure power spectra shown in Fig.~\ref{fig:spectra_slope}a, one can identify three regions of $n$ variations with $\Wi$: before the instability at $\Wi<\Wi_c$, $n\approx-2.2\pm 0.3$; above that, $|n|$ grows with $\Wi$ in the range $\Wi_c < \Wi \lesssim 400$, corresponding to the transition regime; further, $|n|$ grows and reaches up to $3.9\pm0.2$ at $\Wi = 613$, at the ET regime; then, for $\Wi \gtrsim 900$ and up to the highest $\Wi$ value measured, $|n|$ decreases with $\Wi$ down to $|n|\approx 3.2\pm0.2$ at $\Wi=1478$, indicating the DR regime~\cite{Varshney2018, Steinberg2021, Jha2020}. Moreover, the trend observed for the pressure spectra decay slopes, $n$, versus $\Wi$ is similar to the trend for the velocity spectra decay slopes, $m$ versus $\Wi$, though the range of change in the slope tendency is unexpectedly shifted towards the higher $\Wi$ values (Fig.~\ref{fig:spectra_slope}a, and b). Indeed, for $\Wi \lesssim \Wi_c$ the decay exponents are low ($|m|<2$). Above $\Wi_c$, $|m|$ grows in the range $\Wi_c \lesssim \Wi \lesssim 650$, reaching up to $|m|=2\pm 0.2$; at $650\lesssim \Wi\lesssim 1000$ the exponent $|m|$ reaches up to $|m|=2.8 \pm 0.2$ at $\Wi=938$, indicating ET; lastly, for $\Wi\gtrsim 1000$ the exponent $|m|$ reduces down to $|m|=2.2\pm0.2$ at the highest measured Weissenberg value of $\Wi=2049$, indicating the DR regime. Thus, using the decay exponents of the pressure and velocity power spectra, $n$ and $m$, we can divide our data into three regimes above a laminar flow at $\Wi>\Wi_c$: the transition, ET and DR regimes, all of which are characterized by a chaotic flow.

\begin{figure}[h!]
	\centering
	\includegraphics[width=8.58cm]{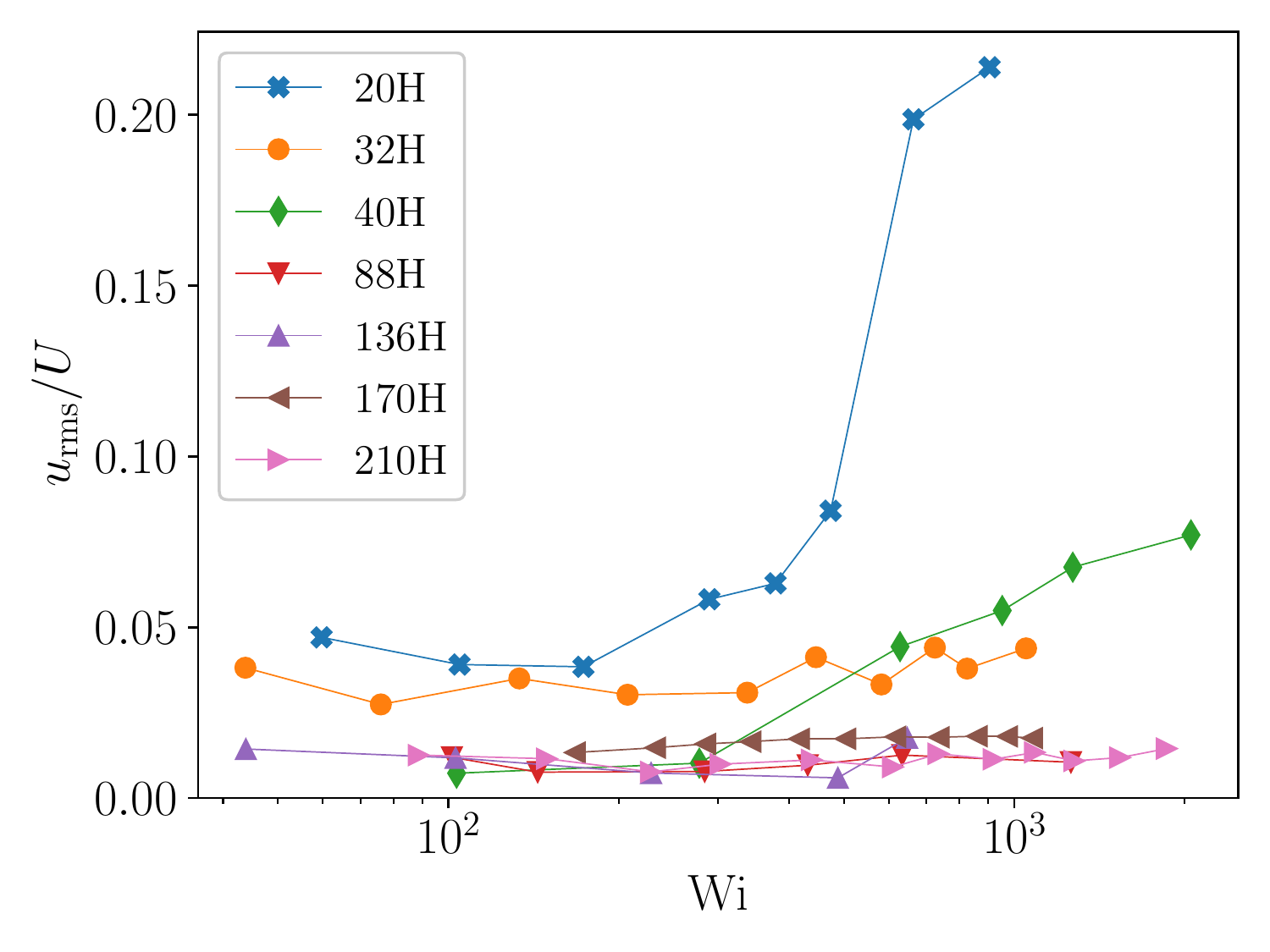}
	\caption{The root mean square of velocity fluctuations, $u_{\mathrm{RMS}}/U$, at the center-line of the channel, $z=0$, shown at various stream-wise locations downstream from the cavity as a function of $\Wi$. 
		\label{fig:uRMS_vs_distance}}
\end{figure}

{\it Coherent structures in three flow regimes.} One of our key observations in the straight channel visco-elastic flow with weak perturbations is stream-wise streaks that occur in the three flow regimes at $\Wi>\Wi_c$. Figure~\ref{fig:40H_structure} presents a series of 12 instantaneous stream-wise velocity fluctuation maps in the x-z central plane and $S=40H$, where the mean velocity profile, $\overline{u}(z)$, was subtracted from the fully measured stream-wise velocity. The sequence of images is shown at with a time step of 0.3 s and for a full duration of 3.3 s at $\Wi=938$ in ET; notably, the time 3.3 s approximately corresponds to the period of elastic waves at this $\Wi$ (f=0.3 Hz) as shown below in Fig.~\ref{fig:40H_structure}. The images demonstrate the occurrence of counter-propagating streaks: span-wise modulated stream-wise velocity fluctuations. Probably, due to low values of elastic wave intensities expected to synchronize a cycle, as found in the channel flow with strong perturbations, and in spite of sufficiently large stream-wise velocity fluctuations, it is impossible to quantitatively verify a cycle period in the streak temporal dynamics, the approach used in Ref.~\cite{Jha2021}. Indeed, at $S=40H$, $u_{\mathrm{RMS}}/U$ increases from about $4.5\%$ at $\Wi=637$ in the transition regime, to about $5.5\%$ at $\Wi=938$ in ET, increases further to about $6.5\%$ at $\Wi=1287$, and it reaches a maximum about $8\%$ at $\Wi=2046$ in DR, comparable with the values observed in Ref.~\cite{Jha2021}, whereas at $40H<S\leq 210H$ and in the same range of $\Wi$ the values one finds for $u_{\mathrm{RMS}}/U$ are less than $2.5\%$ and streaks are not detected (see Fig. \ref{fig:uRMS_vs_distance}). As shown in Fig.~\ref{fig:40H_structure}, the streaks are unsteady, and they appear and disappear seemingly at random. In this sequence, a high-velocity streak (red) emerges and then deteriorates at the channel center in the x-z plane. The random alteration of streaks is further demonstrated in the Supplamentary Materials~\cite{supplamentaryMaterial} through an animation at $\Wi=938$ in ET (see Movie S1), and through three other series of images at $\Wi=637$ in the transition (Fig.S2), at $\Wi=1287$ in DR (Fig.S3), and at $\Wi=2046$ in DR (Fig.S4) as well.

\begin{figure}
	\centering
	\includegraphics[width=8.58cm]{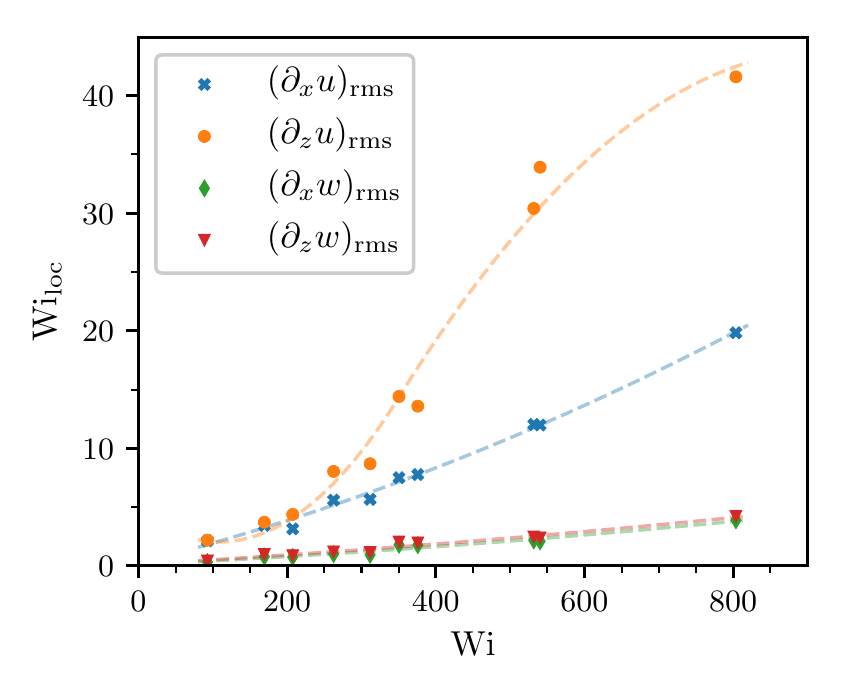}
	\caption{The local Weissenberg number, $\Wi_{\mathrm{loc}}$, is calculated using four components of the velocity gradient tensor in the x-z central plane and plotted as a function of $\Wi$ at $S=80H$.
		\label{fig:wi_loc}}
\end{figure}

{\it Velocity gradient fluctuations and downstream attenuation of velocity fluctuations.} The \textit{local Weissenberg number} is defined as $\Wi_{\mathrm{loc}}\equiv\lambda(\partial u_i/\partial x_j)_{\mathrm{RMS}}$, using the stream-wise velocity gradients fluctuations. As discussed theoretically~\cite{Balkovsky2000} and then demonstrated experimentally~\cite{Liu2010,Liu2014}, in a random flow $\Wi_{\mathrm{loc}}$ defines the degree of polymer stretching and, in particular, the coil-stretch transition taking place at $\Wi_{\mathrm{loc}}^{cst}=1$~\cite{Balkovsky2000,Liu2014}. Only above $\Wi_{\mathrm{loc}}^{cst}$, ET shows up \cite{Steinberg2021}, where polymers are stretched by a chaotic flow almost up to full length \cite{Gerashchenko2005, Liu2010, Liu2014}, particularly at low polymer concentration $c_p$ (see the inset of Fig.~3 in Ref.~\cite{Liu2010} for $c_p=6.29~\mu g/ml$, where relative stretching $x_p/L>0.9$ at $\Wi_{\mathrm{loc}} \sim 30$). In Fig.~\ref{fig:wi_loc}, we plot $\Wi_{\mathrm{loc}}$ as a function of $\Wi$ at $S=80H$. The figure presents $\Wi_{\mathrm{loc}}$ calculated for four available components of the velocity gradient tensor in the x-z central plane. It demonstrates that the most significant of them is the span-wise gradient of the stream-wise velocity fluctuations, $(\partial u / \partial z)_{\mathrm{RMS}}$. In particular, based on the $(\partial u / \partial z)_{\mathrm{RMS}}$ component, at $\Wi\gtrsim \Wi_c$, $\Wi_{\mathrm{loc}}$ gradually increase and eventually reaches a maximum at $\Wi_{\mathrm{loc}}\approx 40$ corresponding to $\Wi\approx800$. Notably, this differs from a channel flow with strong stream-wise perturbations at the inlet, where $(\partial u / \partial z)_{\mathrm{RMS}}$ and $(\partial u / \partial x)_{\mathrm{RMS}}$ have about the same values \cite{Jha2020}.

The channel flow in the experiment can be divided into three regions based on the stream-wise position downstream from the cavity at $S>0$, whereas upstream of it, at $S<0$, the flow is laminar. 
Our velocity measurements via PIV reveal that the intensity of stream-wise velocity fluctuations, $u_{\mathrm{RMS}}/U$, decay downstream from the cavity with $S$. This is shown in Fig.~\ref{fig:uRMS_vs_distance} via measurements at the channel center-line $z=0$ and plotted versus $\Wi$ for various $S$ values. Despite significant scatter of the data, it is clearly seen that the most intense fluctuations occur close to the cavity, e.g., at $S=20H$, while for distances $S\gtrsim 80H$, the fluctuation intensity is much weaker, decreasing down to roughly $2.5\%$. In this first region, at $0 < S \lesssim 80H$, we observe above the instability onset chaotic fluctuations with a steep decay of the velocity spectrum at high frequencies and low-frequency peaks (i.e., Fig.~\ref{fig:40H_spectra}). Then, further downstream, at $80H < S \lesssim 200H$, the high-frequency fluctuations in the velocity power spectra become much weaker, while elastic waves are still observed as sharp spectral peaks at low frequencies in the stream-wise velocity power spectra. Lastly, at $S\gtrsim 200H$, the fluctuations and the low-frequency peaks decay further and become too weak to be resolved.

\begin{figure}[h]
	\centering
	\includegraphics[width=8.58cm]{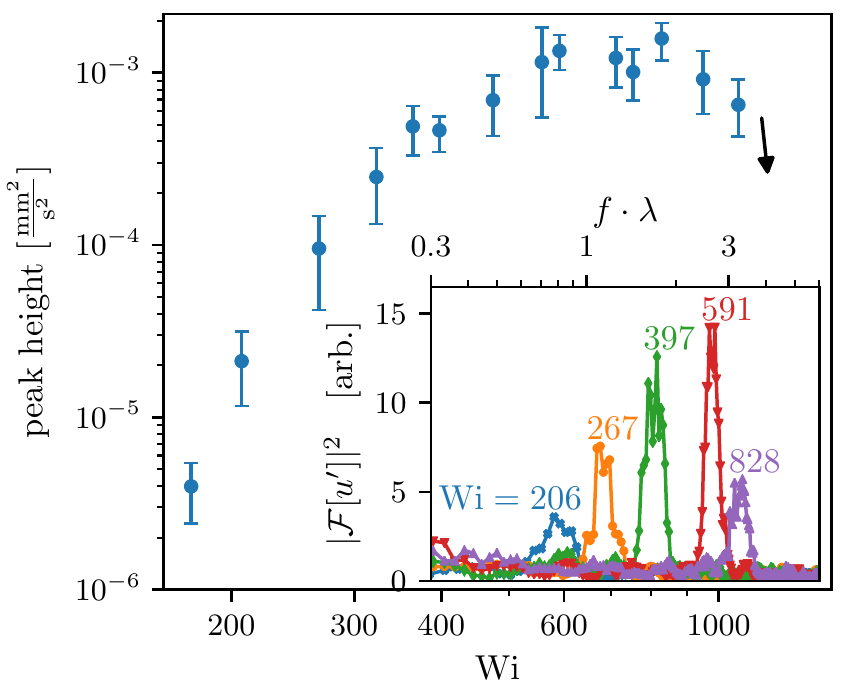}
	\caption{Main panel: height of the spectral peaks observed for the elastic waves, plotted versus $\Wi$ in log-log scales. Inset: stream-wise velocity power spectra at $S=170H$ and $z=\pm\frac{1}{6}H$ for five $\Wi$ values in lin-log scale. The spectra and frequencies are normalized by the noise level and $\lambda$, respectively. \label{fig:u_spectrum}}
\end{figure}

{\it Span-wise propagating elastic waves.} As we discussed above, rather wide noisy peaks of elastic waves at low frequencies in the stream-wise velocity power spectra are observed at downstream locations from the cavity in the range $40H < S < 80H$, similar to those found in~\cite{Jha2020}. However, farther downstream from the cavity, at $S=170H$, we reveal distinct sharp spectral peaks in the stream-wise velocity spectra at a wide range of $\Wi$ values, from above the instability onset and up to roughly $\Wi\approx1000$. The spectral peaks at $S=170H$ are shown in log-linear scales in the inset in Fig.~\ref{fig:u_spectrum} for five $\Wi$ values. The peak's frequency grows with $\Wi$. Furthermore, the normalized peak's intensity as a function of $\Wi$, which is presented in log-log scales in Fig.~\ref{fig:u_spectrum}, grows rapidly from zero at $\Wi_c$ for $170\lesssim\Wi\lesssim 360$; then its growth slows down at $360\lesssim\Wi\lesssim 600$, it saturates at $600<\Wi\lesssim 900$ and drops down to zero by $\Wi\approx 1100$. Moreover, for $\Wi>900$, the peaks become broader and less coherent (not shown). This intensity behavior of the elastic waves agrees well with our earlier observations in the flows between two obstacles and past an obstacle hindering a channel flow, and, particularly, in a straight channel flow with strong perturbations at the inlet ~\cite{Varshney2019,Kumar2021,Jha2020}. In those flow geometries, the dependence of the wave intensity on $\Wi$ correlates with the dependence of the friction factor on $\Wi$ that exhibits the transition, ET, and DR regimes~\cite{Kumar2021,Jha2020}. In the current experiment, the three regions of the $\Wi$ dependence of elastic wave intensity also correlate with the dependence of the decay exponents of the pressure and velocity power spectra shown in Fig.~\ref{fig:spectra_slope}, corresponding to those flow regimes.

\begin{figure}[!htb]
	\centering
	\subfloat{\label{fig:compiled_waves}
		\includegraphics[width=8.58cm]{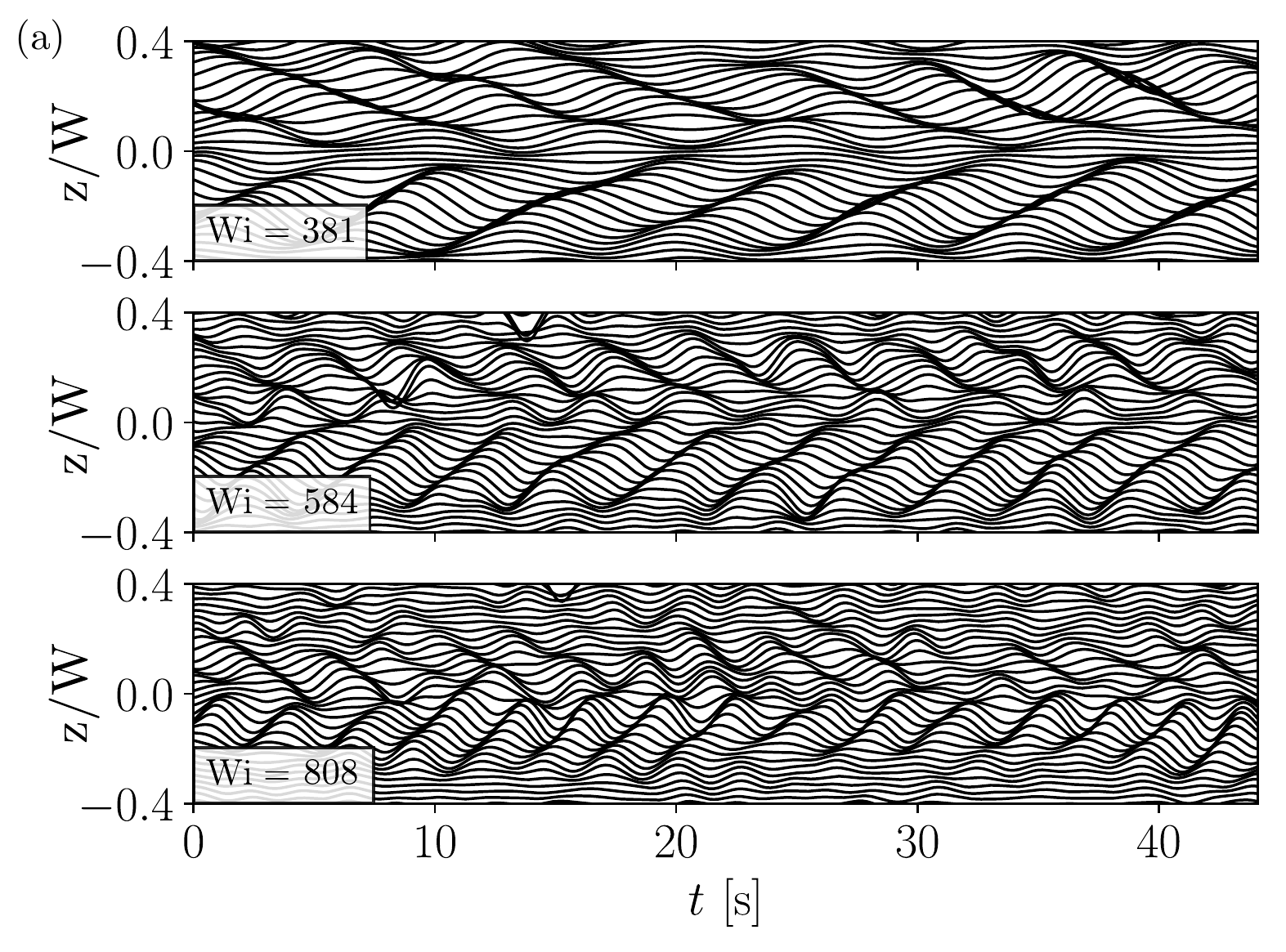}
	}\\[-.3em]
	\subfloat{\label{fig:phase_averaged}
		\includegraphics[width=6.5cm]{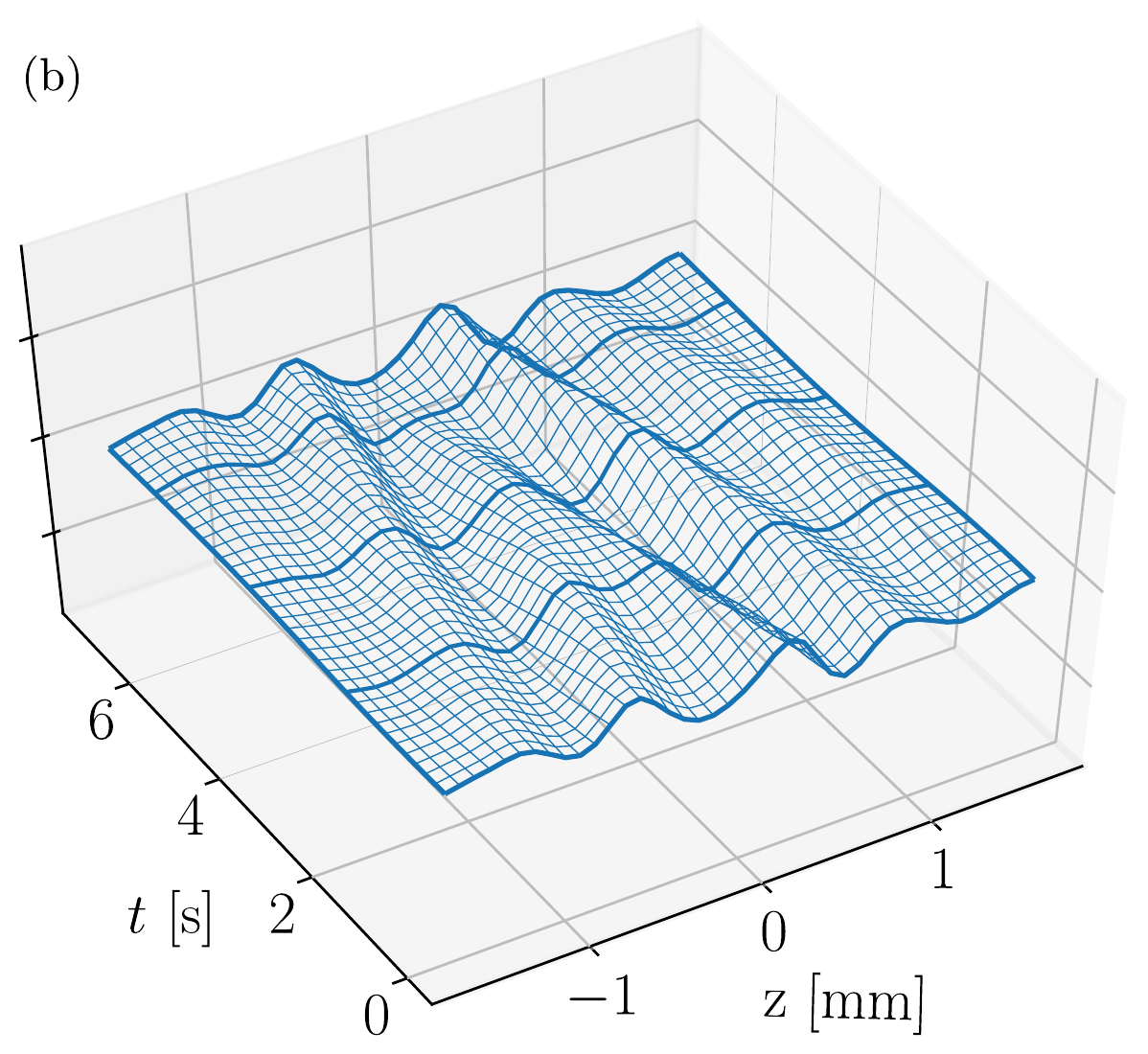}}
	\caption{(a) Space-time plots at $-0.4<\frac{z}{W}<0.4$ of the stream-wise velocity fluctuations, $u'_x(z, t)$ exhibiting elastic wave structures for three values of $\Wi$. The time series are filtered via a band-pass Butterworth filter centered around the spectral peaks to remove background noise. (b) stream-wise velocity, phase averaged at the elastic wave frequency for $\Wi=407$.} \label{fig:wavefoRMS}
\end{figure}

To examine the spatial structure of the velocity field, we plot the stream-wise velocity fluctuations, band-pass filtered around the spectral peak frequency, as space-time plots in Fig.~\ref{fig:compiled_waves}. The structure is also shown more explicitly by phase averaging the velocity fluctuation signals in Fig.~\ref{fig:phase_averaged} for $\Wi=407$. Fig.~\ref{fig:phase_averaged} reveals the wavy structure of the velocity fluctuations propagating in the span-wise direction that is very well described by the expression 
\begin{equation}
u'_x(t,z) = A\,\exp 2 \pi i (f\,t + l^{-1} |z|) \,\, .
\label{eq:waveform}
\end{equation}
Here $A$, $f$ and $l$ are the amplitude, frequency, and wavelength, respectively. Thus, the structure in Fig.~\ref{fig:wavefoRMS} corresponds to transverse span-wise propagating waves. Furthermore, the angle between the wave crests and the horizontal direction in Fig.~\ref{fig:compiled_waves} corresponds to the wave velocity, $v$. Thus, from changes of the angle with $\Wi$, seen in Fig.~\ref{fig:compiled_waves}, we obtain the dependence of the wave velocity on $\Wi$. Notably, one of the most distinctive features of the elastic waves reported here turns out to be the direction of elastic waves propagation: while here the propagation is in the span-wise direction, in all other flow geometries discussed in Refs.~\cite{Varshney2019,Kumar2021,Jha2020} the propagation is in the stream-wise direction.

\begin{figure}[!htb]
	\centering
	\includegraphics[width=8.6cm]{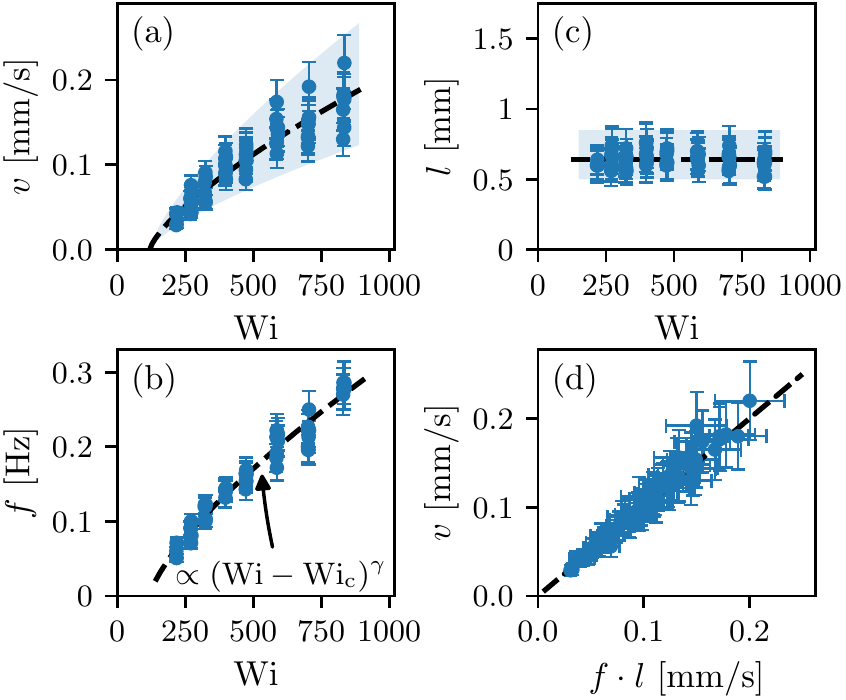}
	\caption{(a), (b) and (c) show the wave velocity, frequency, and wavelength dependence on $\Wi$, respectively. Dashed lines are the best fits. (d) shows the wave velocity versus $f\cdot l$ that confiRMS the linear dispersion relation shown as a dashed line.}
	\label{fig:wave_parameters}
\end{figure}

In Fig.~\ref{fig:wave_parameters}a,b,c, we plot the three main wave characteristics, $f$, $l$, and $v$, as functions of $\Wi$. The data reveals that $l$ does not depend on $\Wi$ in the range explored in the experiments, although it randomly fluctuates in the range $l \in [0.5,0.8] \, \mathrm{mm} \approx [0.14W, 0.22W]$. It seems that $l$ is determined by the channel width, and in particular, $l$ is about 1/3 of the half channel width. On the other hand, as seen in the inset in Fig.~\ref{fig:u_spectrum}, $f$ grows significantly with $\Wi$; its dependence on $\Wi$ can be fitted by the power-law $f \sim (\Wi-\Wi_c)^{\gamma}$ with $\gamma = 0.73 \pm 0.05$, as shown in Fig.~\ref{fig:wave_parameters}b. Furthermore, in Fig~\ref{fig:wave_parameters}d, we plot $v$ versus $f \cdot l$ that verifies the dispersion relation $v=f \cdot l$ within experimental uncertainty. This confirms the existence of the linear dispersion for the elastic waves for the first time, as predicted in~\cite{Balkovsky2001, Fouxon2003}. Lastly, $v$, obtained from the angle between the wave crests and the horizontal direction for different $\Wi$, is fitted with the same power law as obtained for the frequency dependence $v= A(\Wi - \Wi_c)^\gamma$ with the coefficient $A\simeq 3\times 10^{-3}$ mm/s, shown in Fig.~\ref{fig:wave_parameters}a. It is noticeable that $A$ is about three orders of magnitude smaller than the values found in all other flow geometries, where the stream-wise elastic waves are observed, whereas $\gamma$ is roughly the same~\cite{Varshney2019, Jha2020}.

\section{Discussion and conclusions}

In this paper, we address several questions posted in the Introduction regarding the nature of the elastically driven instability in the straight channel visco-elastic flow at $\Re \ll 1$ and $\Wi \gg 1$. In particular, we tackle the main problem related to the recent observations of the elastic instability and chaotic flow in pipe and square channel shear flows strongly perturbed at the inlet \cite{Bonn2011,Pan2013}. Above the instability onset, in a planar channel visco-elastic flow with strong prearranged perturbations at the inlet, the transition, ET, and DR regimes, and elastic waves were discovered in Ref.~\cite{Jha2020}. Our results reported in the current paper clearly demonstrate that even very weak perturbations, initiated by a small cavity located at the top wall and in the middle of the channel, are capable to excite the elastic instability along with all three chaotic flow regimes at higher $\Wi$. The elastic instability threshold at $\Wi_c=120 \pm15$ is determined from the $\Wi$ dependence of the normalized RMS stream-wise velocity fluctuations $(u_{\mathrm{RMS}}-u_{\mathrm{RMS,lam}})/U$ (Fig. \ref{fig:uRMS_transition}) and the normalized RMS pressure fluctuations $\delta P_{RMS}=P_{RMS}/P_{RMS,lam}-1$ (Fig. \ref{fig:dp_RMS}a). Above the instability onset, both $(u_{\mathrm{RMS}}-u_{\mathrm{RMS,lam}})/U$ and $\delta P_{\mathrm{RMS}}$ grow with $\Wi$ algebraically with exponents of $a=0.85$ and $\alpha=1.5$, respectively. The exponent values significantly differ from $0.5$, the value expected for the linear normal mode bifurcation~\cite{Drazin2004}. Moreover, at $\Wi>\Wi_c$, a randomly fluctuating, chaotic flow is found (see Fig.~\ref{fig:dp_RMS}b and Fig.~\ref{fig:40H_spectra}a,b). These observations indicate that the elastic instability is the same non-normal mode bifurcation that was already observed and characterized in the case of strong perturbations at the inlet~\cite{Jha2020}. Thus, our finding confirms the early predictions of the linear stability of visco-elastic parallel shear flows at $\Re\ll 1$ and $\Wi\gg 1$. It also answers our first and the most important question posed in the Introduction, and confirms that weak but finite-size perturbations can lead to elastic instability and even further to ET and DR at higher $\Wi$.

In addition to this, even prior to the instability, we reveal continuous spectra for the pressure fluctuations (Fig.~\ref{fig:dp_RMS} at $\Wi<\Wi_c$) with the power-law decay at high frequencies, despite the laminar flow expected. This observation probably occurs due to a $1/f$ instrumental noise in the pressure measurements. The reason for finite-size perturbations generated by the cavity leading to the elastic instability at $\Wi>\Wi_c$ is, probably, the huge difference in the critical values of the elastic instability inside the cavity $\Wi_{crit}$ (due to the linear elastic instability inside the cavity~\cite{Pakdel1996, Pakdel1998}) and $\Wi_c$ in a straight channel visco-elastic flow downstream the cavity. Indeed, using Eq.~(4) of Ref.~\cite{Pakdel1996}, one gets that for a cavity of $D=0.5$ mm and height of $h=5$mm $\Wi_{crit}=(\alpha\Lambda+\beta)=4.2$, where $\Lambda=h/D=10$, and $\alpha=0.14$, $\beta=2.8$, $M_{crit}=1$ are the constants taken from~\cite{Pakdel1996}. Thus, $\Wi_{crit}$ of the elastic instability inside the cavity is up to two orders of magnitude less than the $\Wi_c=120$ measured here in the channel. 

%Our measurements of velocity and pressure fluctuations demonstrate that the small cavity produces a weak but finite-size perturbations, which are sufficient to excite an elastic instability at $\Wi > \Wi_c \simeq 120$. Yet

%As the result, at $\Wi$ values close to $\Wi_c$, ET occurred in the cavity engenders a wide spectrum of perturbations in the channel flow downstream of the cavity, whereas upstream of the cavity a laminar flow without measurable random perturbations is found. Moreover, in spite of the fact that also in this case the perturbations are not controlled, one may measure them or perform numerical simulations to getstream-wisevelocity fluctuations as function of $U$, the stream-wise mean velocity of the channel flow.

As discussed in the Introduction, in the strongly perturbed channel flow, three flow regimes were identified at $\Wi>\Wi_c$ over a wide range of $\Wi$ in Refs.~\cite{Jha2021, Jha2020}. In the case of a weakly perturbed channel flow of visco-elastic fluid, the flow regimes at $\Wi>\Wi_c$ are identified similarly to the strongly perturbed one~\cite{Jha2020}, using the $\Wi$ dependence of the following observable: the decay exponents of the pressure and velocity spectra in Fig.~\ref{fig:spectra_slope}a,b, and the intensity of the elastic waves shown in Fig.~\ref{fig:u_spectrum}. From the dependence of the decay exponents' values on $\Wi$, three flow regimes at $\Wi>\Wi_c$ are identified. The exponent absolute values increase up to $|n|\approx 3\pm0.3$ at $\Wi\approx 400$ for the pressure spectrum, and up to $|m|\approx 2\pm0.2$ at $\Wi\approx 650$ for the velocity spectrum. This range of $\Wi_c<\Wi\leq 650$ is defined as the transition regime. Then, both decay exponents grow further: for the pressure it reaches up to $|n|=3.9\pm0.2$ at $\Wi\approx 650$ for the velocity up to $|m|=2.8\pm0.2$ at $\Wi=960$ that define the ET regime for pressure and velocity spectra, respectively. Further, in the DR regime, both exponents decrease: for the pressure spectrum down to $|n|=3.3\pm0.2$ at $\Wi=1478$ and for the velocity spectrum down to $|m|=2.2\pm0.2$ at $\Wi\approx2049$. In addition to that, three flow regimes can be identified in the $\Wi$ dependence of the elastic wave intensity as well. At $\Wi_c<\Wi\lesssim 650$ in the transition regime, the elastic wave intensity increases; for higher Weissenberg number values it reaches a narrow plateau which identified as ET at $650 \lesssim \Wi \lesssim  900$, and at $\Wi>900$, the elastic wave intensity decays in the DR regime.

Despite the weak perturbations, the channel flow also exhibits CS in the form of counter-propagating streaks in the frame moving with the averaged velocity profile, $\overline{u}(z)$. The streaks are detected in the channel flow only up to $S=40H$ in the three chaotic flow regimes: transition at $\Wi=637$ (Fig.~S2), ET at $\Wi=938$ (movie~S1), and DR at $\Wi=1278$ and $\Wi=2046$ in (Fig.~S3 and Fig.~S4). It turns out that up to $S=40H$ the velocity fluctuations $u_{\mathrm{RMS}}/U$ (Figs.~\ref{fig:profiles}, and \ref{fig:uRMS_transition}) are sufficiently strong to become self-organized into CS. Indeed, at $S=40H$, $u_{\mathrm{RMS}}/U$ reaches up to $\approx 4.5\%$ at $\Wi=637$ in the transition regime, $\approx 6.5\%$ at $\Wi=1280$, and $\approx 8\%$ at $\Wi=2046$ in DR (Figs.~\ref{fig:profiles}). These fluctuation levels are comparable with those detected in a channel flow strongly perturbed at the inlet, where CSs are observed and studied at $\Wi=1050$ in a wide range of $l/h$. At $40h\leq l \leq 140h$, $u_{\mathrm{RMS}}/U$ reaches $8-10\%$ and decreases down to $\approx4\%$ in the range $150h<l\simeq 220h$, where the CSs disappear \cite{Jha2020}. Thus, the value of the velocity fluctuations defined the appearance and existence of CSs. However, the streak temporal dynamics is not synchronized into a cycle by the elastic waves, as in a channel flow strongly perturbed at the inlet, where the CS cycling period precisely coincides with that of the elastic waves \cite{Jha2020}. This can be explained by their significantly lower intensity, which is not an ample energy source to organize CS, and their role in the streak existence is minimal. Moreover, since the flow structure with streaks changes continuously, it is hard to verify the period of the transformations, though the streaks are clearly identified. Notably, the appearance of CSs, discussed in the transient growth theory of Refs.~\cite{Jovanovic2010, Jovanovic2011, Page2014}, agrees with the non-modal instability interpretation and appearance of CSs, presented in this paper.

Unlike the previous observations of elastic waves in other flow configurations~\cite{Varshney2019, Jha2020, Kumar2021}, the elastic waves in our experiment are propagating in the span-wise direction. We can explain this observation by the values of the RMS fluctuations of the different components of the velocity gradients tensor, which defines the degree of polymer stretching, and so the elastic stress tensor components (Fig.~\ref{fig:wi_loc}). Indeed, at $\Wi\geq \Wi_c$, the local Weissenberg number, $\Wi_{\mathrm{loc}}\equiv\lambda(\partial u_i/\partial x_j)_{\mathrm{RMS}}$, gradually increases with $\Wi$ for all tensor components, but the largest value is obtained for the $(\partial u/\partial z)_{\mathrm{RMS}}$ component, being $\Wi_{\mathrm{loc}}\approx 40$ at $\Wi\approx 800$. Thus, the elastic stress is the highest for the gradient in the span-wise direction of the stream-wise velocity. The observation of the span-wise propagating elastic wave suggests that its propagation direction is defined by the largest RMS velocity gradient $(\partial u/\partial z)_{\mathrm{RMS}}\gg (\partial u/\partial x)_{\mathrm{RMS}}$ and not by the mean stream-wise velocity of a channel shear flow. In contrast, in the straight channel flow with strong perturbations at the inlet, the components $(\partial u/\partial x)_{\mathrm{RMS}}$ and $(\partial u/\partial z)_{\mathrm{RMS}}$ are of about the same value, and the elastic waves propagate in the stream-wise direction, probably, due to additional small polymer stretching in the shear flow direction \cite{Jha2020}.

The strong and chaotic fluctuations of the flow decay up to $\sim 40H$ downstream the cavity, whereas weak but coherent span-wise elastic waves could be identified up to $S=170H$ due to two factors. First, the RMS of velocity fluctuations decays and become significantly smaller than the elastic wave intensity at this location. Second, due to very small frequency of the elastic waves, their attenuation is extremely small and many orders of magnitude lower than in the strongly perturbed at the inlet channel flow \cite{Jha2020}. At $S=170H$, we measure the dependence of the wave's frequency, wavelength, and velocity on $\Wi$ independently, and thus we confirm the theoretical predictions on their linear dispersion relation~\cite{Balkovsky2001,Fouxon2003}. Furthermore, the dependence of the parameters of elastic waves on $\Wi$ is similar to that found previously in other flow geometries \cite{Varshney2019,Jha2020}. Nevertheless, in spite of the same value of the scaling exponent $\gamma$ in $v = A(\Wi - \Wi_c)^\gamma$, the coefficient we find here, $A\simeq 3\times 10^{-3}$ mm/s, is almost three orders of magnitude smaller than the value $A\simeq 0.5$ mm/s obtained in the previous measurements of the stream-wise propagating elastic waves~\cite{Varshney2019, Jha2020}. To explain the lower velocity values in the weakly perturbed channel flow, we recall that the velocity of elastic waves depends on the magnitude of the elastic stress in the in the flow~\cite{Balkovsky2001, Fouxon2003}. Since for the span-wise propagating elastic wave the velocity is defined by the span-wise elastic stress component, the low velocity means that the low elastic stress that is confirmed by about 25 times difference between the maximum value of $\Wi_{\mathrm{loc}}\approx 40$ in weakly perturbed flow Fig.~\ref{fig:wi_loc} versus $\Wi_{\mathrm{loc}}\approx 1020$ in the strongly perturbed channel flow \cite{Jha2020}. Nevertheless, it is not obvious, why $A$ should be so different from its value for the stream-wise propagating waves.

To summarize, let us return to the questions posed in the introduction. (i) We find that the strong perturbations are not necessary to get elastic instabilities in straight channel flows of visco-elastic fluids at $\Re\ll 1$ and $\Wi\gg 1$. (ii) The growth of velocity and pressure fluctuations in Figs.~\ref{fig:uRMS_transition} and \ref{fig:dp_RMS}, as well as the continuous spectra above the instability onset in Fig.~\ref{fig:40H_spectra} indicate that the transition occurs due to a non-modal bifurcation. (iii) The trends we observe for the velocity and pressure spectra decay exponents in Fig~\ref{fig:spectra_slope}, as well as for the elastic wave intensity dependence on $\Wi$ in Fig.~\ref{fig:u_spectrum}, suggest the existence of three flow regimes, similarly to those characterized in Ref.~\cite{Jha2021, Jha2020}. Furthermore, we find CS, namely streaks, which are detected in three flow regimes at $S=40H$. At larger $S$, the RMS velocity fluctuations become too small to be self-organized into CS. (iv) We have observed elastic waves in three flow regimes, and confirmed in Fig.~\ref{fig:wave_parameters} their linear dispersion-relation for the first time. (v) We observe a correlation between the elastic wave intensity and the pressure and velocity spectra decay exponents; this could hint on a common mechanism underlying the two phenomena and agrees with the suggestion of Refs.~\cite{Jha2021, Jha2020} regarding the elastic wave's energizing role in the self-sustained process. Thus, it appears that the nature of the non-normal mode instability, the existence of the three flow regimes, and the existence of elastic waves, do not depend on the perturbation strength, although the details of the flow, such as the intensity of the flow fluctuations, the propagation direction of elastic waves and extremely low values of their velocity, are sensitive to the perturbations. The observed similarity in flows with weak and strong perturbations may hint on universality in the development of three flow regimes in elastically driven visco-elastic channel flow, independent of amplitudes of finite perturbations. There is also limited similarity in the appearance CSs in Newtonian turbulent channel flows, though the mechanisms of CSs generation are drastically different \cite{Grossmann2000,Schoppa2002}. This important issue requires further study in other flow geometries.

\begin{acknowledgements}
	
	We are grateful to Guy Han, Rostyslav Baron, and Gershon Elazar for their assistance with preparing the experimental setup. This work was partially supported by grants from the Israel Science Foundation (ISF; grant \#882/15 and grant \#784/19) and the Binational USA-Israel Foundation (BSF; grant \#2016145). RS is grateful for the financial support provided by the Clore Israel Foundation.
	
\end{acknowledgements}

\bibliography{bib}

%\clearpage

\end{document}